\documentclass[amsmath,amssymb,showpacs,floatfix,nofootinbib,showkeys,prl]{revtex4}

\usepackage{graphicx}% Include figure files
\usepackage{dcolumn}% Align table columns on decimal point
\usepackage{bm}% bold math
\usepackage{color}
\usepackage{extarrows}
\usepackage{enumitem}
\usepackage{threeparttable}
\usepackage{relsize} % make math symbols larger
\def\urlprefix{} \def\url#1{}
 \def\isbn#1{}
\def\eprint#1{}
\newcommand{\Gl}{Eq.}
\newcommand{\gl}{Eq.}

\newcommand{\gls}{Eqs.}

\graphicspath{{../digamma/}{../Fe-CnDT_Nijhuis/}{.}}
\DeclareGraphicsExtensions{.eps,.jpg,.tif,.pdf,.png}
\usepackage{hyperref}
\newcommand{\ib}[1]{{\color{black}#1}}
\newcommand{\corr}[1]{{\color{blue}#1}}

\newcommand{\figname}{Fig.~} 
\newcommand{\figsname}{Figs.~}
\begin{document}

\title[Significant Insight]{
% 
  %  Exact formula for the tunneling conductance: A unified framework for Arrhenius and Sommerfeld thermal regimes also revealing important differences between nanojunctions with single molecule architectures versus self assembled monolyers
  %  Exact formula for the tunneling conductance spanning Arrhenius' and Sommerfeld's thermal regimes, and revealing important differences between nanojunctions with single molecule architectures versus self assembled monolyers
% 
  Exact analytic formula for conductance predicting a tunable Sommerfeld-Arrhenius thermal transition within a % overall
  single-step tunneling mechanism in molecular junctions subject to mechanical stretching
}
\author{Ioan B\^aldea\homepage{https://www.pci.uni-heidelberg.de/tc/usr/ioan/}}
\affiliation{%
Theoretical Chemistry, Heidelberg University, Im Neuenheimer Feld 229, 
D-69120 Heidelberg, Germany}

\begin{abstract}
  We show that the conductance $G$ of molecular tunnel junctions wherein the charge transport
  is dominated by a single energy level can be expressed in closed analytic form which is
  exact and valid at arbitrary temperature $T$ and model parameter values. On this basis,
  we show that the single-step tunneling mechanism is compatible with a continuous thermal
  transition from a weakly $T$-dependent $G$ at low $T$ (Sommerfeld regime)
  to a nearly exponential $1/T$-dependent $G$ at high $T$ (Arrhenius-like regime).
  We predict that this Sommerfeld-Arrhenius transition can be observed in real molecular junctions % (e.g., based on perylene diimide)
  and can be continuously tuned, e.g., via mechanical stretching.
\end{abstract}
\keywords{molecular junctions, charge transport, electron tunneling, conductance, thermal effects}
\maketitle
\section{Introduction}
\label{sec:intro}
As a rule, temperature independent conductance data measured in molecular junctions are 
interpreted as evidence
of single-step charge transport by tunneling while a pronounced temperature dependent conductance is taken
as indication of a two-step hopping mechanism
\cite{Frisbie:07,Choi:08,Choi:10,Choi:10b,Frisbie:10,Tao:10,Frisbie:11b,Frisbie:11c,Frisbie:14c,Frisbie:15,Frisbie:16a,Frisbie:16c,Guo:17}.
Notable
exceptions claiming a significant $T$ dependence of the transport by tunneling exist
\cite{Poot:06,Tao:07c,Lambert:11,McCreery:13b,Lewis:13,Tao:16b,McCreery:16a,Nijhuis:16b,Baldea:2017g,Baldea:2018a}.
\ib{In nearly resonant cases ($\left\vert \varepsilon_0 \right\vert \alt 0.1$\,eV), temperature dependent currents can arise due to 
  the thermal broadening of the electronic Fermi distribution in electrodes \cite{Poot:06,Nijhuis:16b}.
  Where experimentally available, the impact of the electrodes' work function turned out to be extremely useful
  for discriminating whether the temperature dependence is due to transport by hopping or transport by tunneling \cite{Baldea:2018a}.
Still,} a thorough investigation to clarify this issue has not been attempted, and this is one of the main aims of the present work.

In our analysis, we will focus on the low bias conductance $G$ of a molecular junction wherein the charge transport
is dominated by a single level (molecular orbital, MO).
We report an analytic formula for $G$ which is exact and valid for arbitrary values
of all relevant parameters (temperature $T$, MO-electrode couplings $\Gamma$'s, and MO energy offset
$\varepsilon_0$ relative to electrodes' Fermi energy).

Based on this exact expression, we are able to specify the parameter ranges
wherein the transport by tunneling yields a $G$ strongly dependent on $T$.
Pleasantly, these ranges turn out to comprise values characterizing real molecular junctions.

The presently deduced formula for $G$ predicts the possibility that, without changing the 
single-step transport mechanism, upon rising the temperature, 
a tunneling junction can continuously switch from
 a weakly $T$-dependent $G$ at low $T$ (Sommerfeld regime)
to a nearly exponential 
$1/T$-dependent $G$ at high $T$ (Arrhenius-like regime). Furthermore, it predicts that in real junctions
the location of this
Sommerfeld-Arrhenius transition can be continuously tuned
because $\Gamma$ and/or $\varepsilon_0$ can be controlled by a mechanical stretching force 
and/or via molecular orbital gating, respectively.

\section{Model and General Formulas}
\label{sec:model}
Applied to uncorrelated transport, the general Keldysh formalism \cite{Caroli:71a,Meir:92,CuevasScheer:17} 
yields the following expression of the
low bias conductance $ G \equiv  \left . \partial I(V)\partial V\right\vert_{V\to 0} $
of a single molecule tunneling junction at finite temperatures $T = 1/\left(k_B \beta \right)$ \cite{Baldea:2017d}
\begin{eqnarray}
  \frac{G}{G_0} & = & - \int_{-\infty}^{+\infty} d\,\varepsilon \, \mathcal{T}\left(\varepsilon \right) \frac{\partial}{\partial \varepsilon} f(\varepsilon) \nonumber \\
  & = & \frac{\beta \Gamma_{g}^{2}}{4} \int_{-\infty}^{+\infty} \frac{ d\,\varepsilon}{\left(\varepsilon - \varepsilon_0\right)^2 + \Gamma_{a}^2}\, \mbox{sech}^{2}\, \frac{\beta \varepsilon}{2} 
  \label{eq-g}
\end{eqnarray}
Here $G_0 \equiv 2 e^2/h = 77.48\,\mu$S, $\mathcal{T}(\varepsilon)$, and $f(\varepsilon) = 1/\left(1 + e^{\beta \varepsilon}\right)$
are the quantum conductance, transmission function, and Fermi distribution, respectively, and energies are measured relative to
electrodes' Fermi energy ($E_F \equiv 0$).
\Gl~(\ref{eq-g}) assumes a single level (MO) of energy $\varepsilon_0$ 
whose coupling to two flat wide band $s$(ubstrate) and $t$(ip) electrodes 
is quantified by an effective coupling $\Gamma_{g}$
expressed in terms of the individual strengths $\Gamma_{s}$ and $\Gamma_{t}$
\begin{subequations}
\begin{equation}
\Gamma_{g}^{2} = \Gamma_{s} \Gamma_{t} \propto \tau_{s}^{2} \tau_{t}^{2}
\label{eq-Gamma}
\end{equation}
which also give rise
to a smeared Lorentzian-shaped transmission of finite half-width $\Gamma_{a}$ \cite{CuevasScheer:17,Baldea:2021b,Baldea:2022b}
\begin{equation}
\label{eq-Delta}
\Gamma_{a} = \left( \Gamma_{s} + \Gamma_{t}\right) / 2
\end{equation}
\end{subequations}
In general, the MO-electrode exchange integrals are different ($\tau_{s} \neq \tau_{t}$),
which make the geometric and arithmetic averages different from each other ($\Gamma_{g} \leq \Gamma_{a}$). 

Integration by parts of \gl~(\ref{eq-g}) followed by contour integration
yields 
the following general formula for the conductance in terms of the real part of the trigamma function
$\psi^{\prime}(z)$ of complex argument
\begin{equation}
  \label{eq-g-exact}
\frac{G}{G_0} = 
\frac{\Gamma_{g}^{2}}{2 \pi \Gamma_{a} k_B T} \mbox{Re}\, \psi^{\prime} \left(\frac{1}{2} + \frac{\Gamma_{a} + i\, \varepsilon_0}{2\pi k_B T}\right) 
\end{equation}
Recall that the trigamma function
is defined as the derivative of the digamma function,
$\psi^{\prime}(z) \equiv \frac{d}{d\,z}\psi(z)$,
which, in turn, is the logarithmic derivative of Euler's gamma function
\cite{AbramowitzStegun:64}.

Noteworthily, 
\Gl~(\ref{eq-g-exact}) is an exact result valid at arbitrary values of all parameters ($\varepsilon_0$, $\Gamma_{g}$, $\Gamma_{a}$, and $T$)
The well known low temperature limit ($k_B T \ll \Gamma_{a}$) \cite{CuevasScheer:17,Baldea:2012g}
\begin{equation}
  \frac{G}{G_0} \xlongrightarrow{k_B T \ll \Gamma_{a}} \frac{G_{0K}}{G_0}
  = \frac{\Gamma_{g}^{2}}{\varepsilon_0^2 + \Gamma_{a}^2} \xlongrightarrow{\Gamma_{a} \ll \vert \varepsilon_0 \vert}
  \frac{\Gamma_{g}^{2}}{\varepsilon_0^2}
  \label{eq-g0K}
\end{equation}
follows from \gl~(\ref{eq-g-exact}) by virtue of the asymptotic trigamma's expansion % ($z \to  \infty$, $\vert \arg z \vert < \pi$)
\cite{AbramowitzStegun:64}. Above, the zero temperature conductance $G_{0K} \equiv G(T)\vert_{T = 0}$
should not be confused with the conductance quantum $G_0$.

In the opposite limit ($\Gamma_{a} \ll \pi k_B T$, $\mbox{Re}\, z \to 1/2$),
the RHS of \gl~(\ref{eq-g-exact}) reduces to
\begin{subequations}
\label{eq-g-arrhenius}
  \begin{equation}
\displaystyle
\label{eq-g-sech} % \tag{13a}
\frac{G}{G_0} \xlongequal{\Gamma_{a} \ll \pi k_B T} 
\frac{\pi}{4} \frac{\Gamma_{g}^{2}}{\Gamma_{a} k_B T} \mbox{sech}^2 \frac{\varepsilon_{0}}{2 k_B T} 
\end{equation}
which in a more particular limit acquires an Arrhenius-like form
\begin{equation}
  \displaystyle
  \label{eq-g-exp} % \tag{13b}
  \frac{G}{G_0} \xlongequal{\Gamma_{a} \ll \pi k_B T \ll \left\vert \varepsilon_0\right\vert }
  % \frac{G_{p.A}}{G_0} =
  \frac{\pi\Gamma_{g}^{2}}{\Gamma_{a} k_B T} 
  \exp\left(-\frac{\left\vert\varepsilon_0\right\vert}{k_B T}\right)
\end{equation}
\end{subequations}
The particular results of \gls~(\ref{eq-g-sech}) and (\ref{eq-g-exp}) were first reported in refs.~\citenum{Baldea:2017d}
and \citenum{Lambert:11}, respectively.
\Gl~(\ref{eq-g-sech}) is a limiting case of a more general approximate interpolation formula
deduced earlier \cite{Baldea:2017d} for $\Gamma_{s,t} = \Gamma_{g,a} $ 
and parameter values where the peaks of $\mathcal{T}(\varepsilon)$ and $- \partial f(\varepsilon)/\partial \varepsilon$
are sufficiently narrow and well separated of each other \cite{Baldea:2017d}. 

Although exact and general, \gl~(\ref{eq-g-exact}) 
may pose certain practical problems for experimental data processing.
Special functions
(read trigamma function $\psi^{\prime}(z)$) having furthermore a complex argument are not usually implemented
in data fitting softwares routinely employed by experimentalists.

Attempting to meet the legitimate experimentalists' desire of having a formula merely containing
elementary functions of real arguments, we arrived at the following very accurate approximation
($y \equiv \varepsilon_0 / \left(2 \pi k_B T\right)$)
\begin{subequations}
  \label{eq-g-approx} 
  \begin{eqnarray}
  \displaystyle
  \frac{G}{G_0} & \simeq &
  \frac{\pi \Gamma_{g}^{2}}{\corr{4} \Gamma_{a} k_B T} \mbox{sech}^2 \pi y
  + \left(\frac{\Gamma_{g}}{2 \pi k_B T}\right)^2
  \varphi(y)
 \label{eq-g-new} \\
 y & \equiv & \frac{\varepsilon_0}{2 \pi k_B T} \nonumber % \label{eq-y}
 \\
  \varphi(y) & = & \frac{y^2 - 34.7298}{\left(y^2 + 2.64796\right)^2} +
  37.262 \frac{y^2 + 1.12874}{\left(y^2 +  2.17786\right)^3} \nonumber \\
  & & + 3.01373 \frac{y^2 -0.082815}{\left(y^2 + 0.25014\right)^3}
  % \xlongequal{y \gg 1} \frac{1}{y^2} + \frac{1}{4 y^4} % + \mathcal{O}\left(\frac{1}{y^{6}}\right)
% 
  \label{eq-phi} 
\end{eqnarray}
\end{subequations}
\corr{Notice the factor 4 entering denominator of the first term in the RHS of \gl~(\ref{eq-g-new}); it replaces the wrong
  factor 16 (a typo) entering \gl~6 of ref.~\citenum{Baldea:2022c}.}
Thorough tests in broad parameter ranges of experimental interest revealed
that \Gl~(\ref{eq-g-approx})
--- which embodies  insight gained from series expansions, asymptotic behavior,
and recurrence properties of the trigamma function ---
reproduces remarkably well the conductance computed
exactly using \gl~(\ref{eq-g-exact}).
To exemplify, suffice it to say that, in all the figures presented below,
the curves generated via \gls~(\ref{eq-g-exact}) and (\ref{eq-g-approx})
cannot be distinguished from each other within the drawing accuracy.
\section{Results}
\label{sec:results}
It is evident that \gl~(\ref{eq-g-exact}) predicts a tunneling conductance dependent on temperature.
Still, the practically relevant question is whether the predicted dependence on $T$
is sufficiently strong for inferring a single-step
tunneling mechanism based on transport data measured
in real molecular junctions
\cite{Poot:06,Tao:07c,Lambert:11,McCreery:13b,Lewis:13,Tao:16b,McCreery:16a,Nijhuis:16b,Baldea:2018a}.

\begin{figure} % {hbtp}
  \centerline{\includegraphics[width=0.45\textwidth,angle=0]{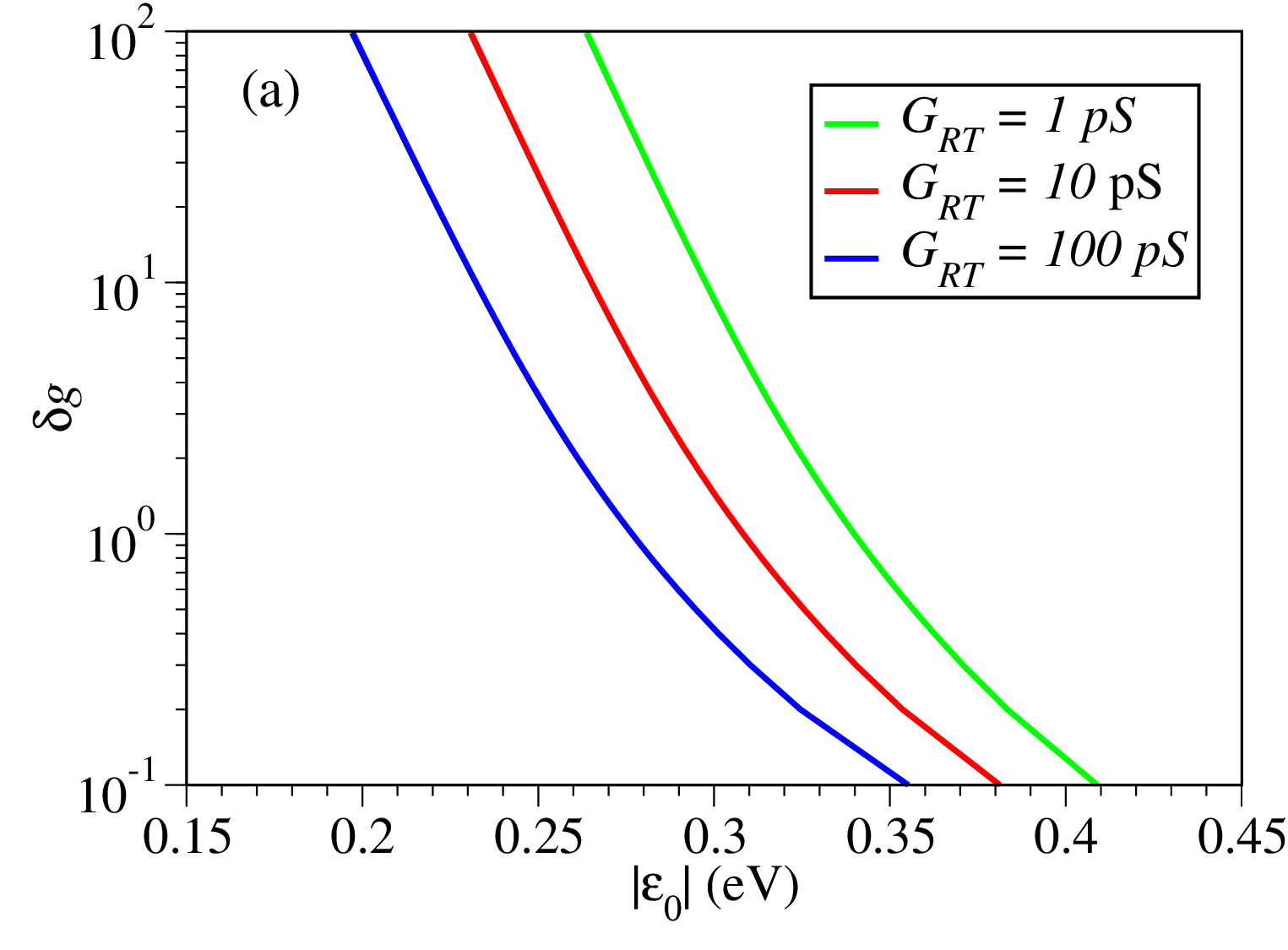}
  } $ $\\[2ex] \centerline{
    \includegraphics[width=0.45\textwidth,angle=0]{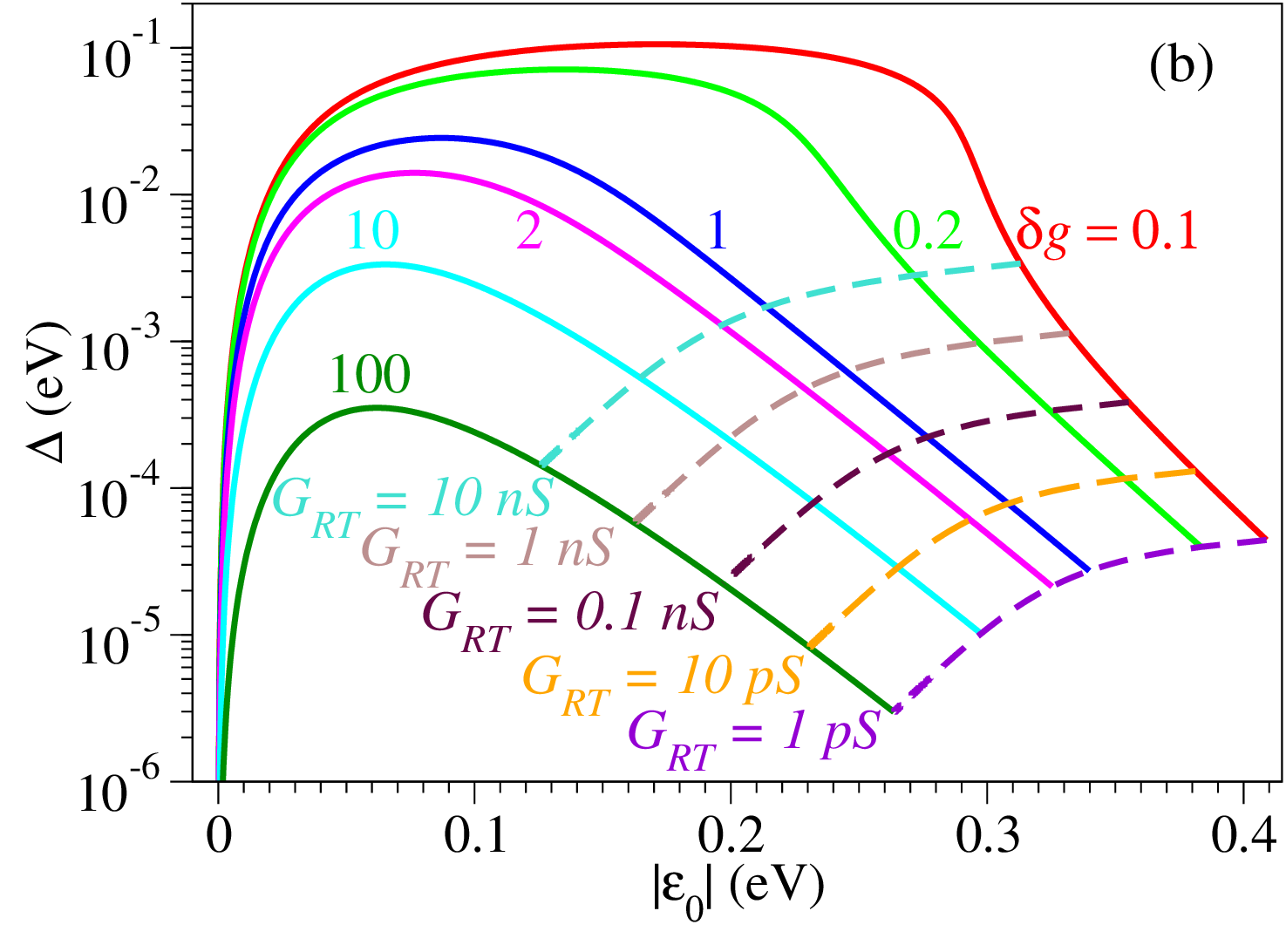}
    % } $ $\\[2ex] \centerline{
    % \includegraphics[width=0.3\textwidth,angle=0]{fig_G_at_DeltaSignificant_RT_various_fracG}
  }
  \caption{(a) Conductance thermal excess at room temperature (RT) relative to zero temperature
    $\delta g \equiv G_{RT}/G_{0K} - 1$ as a function of the MO energy offset
    $\left\vert \varepsilon_0\right\vert$ for several values of the RT conductance $G_{RT}$.
    (b) Curves for the transmission halfwidth $\Gamma_{a}$ versus $\varepsilon_0$ for a given
    value of $\delta g$ (ranging from 0.1 to 100)
    indicated for each curve. Dashed lines correspond to several values of $G_{RT}$ computed
    by assuming $\Gamma_{g} = \Gamma_{a} \equiv \Gamma $.
    }
  \label{fig:significant}
\end{figure}

To answer this question, we used \gl~(\ref{eq-g-exact}) to compute the conductance excess $\delta g \equiv G_{RT}/G_{0K} - 1 $
at room temperature (RT, $T_{RT} = 298.15$\,K) relative to the zero temperature limit.
Results of these numerical calculations are depicted in \figname~\ref{fig:significant}.
To facilitate understanding \figname~\ref{fig:significant}, we note the following. There is no practically relevant
dependence of $G$ on $T$ for molecular junctions (as experimentally
demonstrated in junctions based on alkanes \cite{Reed:09,Zandvliet:12})
with MO offsets $\left\vert\varepsilon_0\right\vert \sim 1$\,eV \cite{Baldea:2019h} much larger
than thermal energies $\sim k_B T_{RT} = 25.7$\,meV. \Gl~(\ref{eq-g-exact}) confirms that
in such cases the conductance thermal excess is altogether negligible $\delta g \simeq 0$.
The smallest value ($\delta g = 0.1  \Leftrightarrow G_{RT}/G_{0K} = 1.1 $)
on the y-axis in \figname~\ref{fig:significant}a corresponds to the lowest excess that can still be
distinguished from very good statistical variances ($\sim 10$\%) inherently present in experiment \cite{Baldea:2017e}.
Depending on the value of $G_{RT} = 1; 10; 100$\,pS,
this conductance excess corresponding to an MO offset
$\left\vert\varepsilon_0\right\vert = 0.409; 0.381; 0.355$\,eV is depicted by the rightmost point on the green, red, and blue curves
of \figname~\ref{fig:significant}a, respectively. Higher conductance thermal excess requires smaller MO offsets.
At the other extreme  value ($\delta g = 99 \Leftrightarrow G_{RT}/G_{0K} = 100$) in \figname~\ref{fig:significant}a,
the MO offsets amounting to $\left\vert\varepsilon_0\right\vert = 0.264; 0.231; 0.197$\,eV
correspond to the leftmost points on the green, red, and blue curves, respectively.

MO energy offsets estimated for real molecular junctions fabricated with Zn-porphyrin \cite{Lambert:11,Lewis:13},
diarylethene \cite{Guo:16a,Baldea:2017g}, and perylene diimide \cite{Baldea:2018a} fall
in the above range ($\left\vert\varepsilon_0\right\vert \approx 0.15 - 0.35$\,eV).
Even smaller values $\left\vert\varepsilon_0\right\vert \approx 50$\,meV were estimated for
alkanedithiolates functionalized with ferrocene \cite{Nijhuis:16b,Baldea:2017d}.
For all aforementioned cases the transport data were found to be temperature dependent.
For obvious physical reasons, a significant dependence $G = G(T)$ implies a sufficiently narrow
transmission peak (i.e., a small $\Gamma_a$). Curves for $\Gamma_a$ versus $\varepsilon_0$
at given values of the RT conductance excess are depicted in \figname~\ref{fig:significant}b.

The important message conveyed by \figname~\ref{fig:significant} should be clear:
orders of magnitude conductance enhancement
upon increasing temperature is fully compatible with the single-step
tunneling mechanism, backing the suggestions put forward in the experimental studies cited above.

\begin{figure} % {hbtp}
  % g_vs_T_Ag_parameters_Manuscript_w_pccp.dat
  % g_vs_T_Ag_parameters_Manuscript.dat
  \centerline{\includegraphics[width=0.45\textwidth,angle=0]{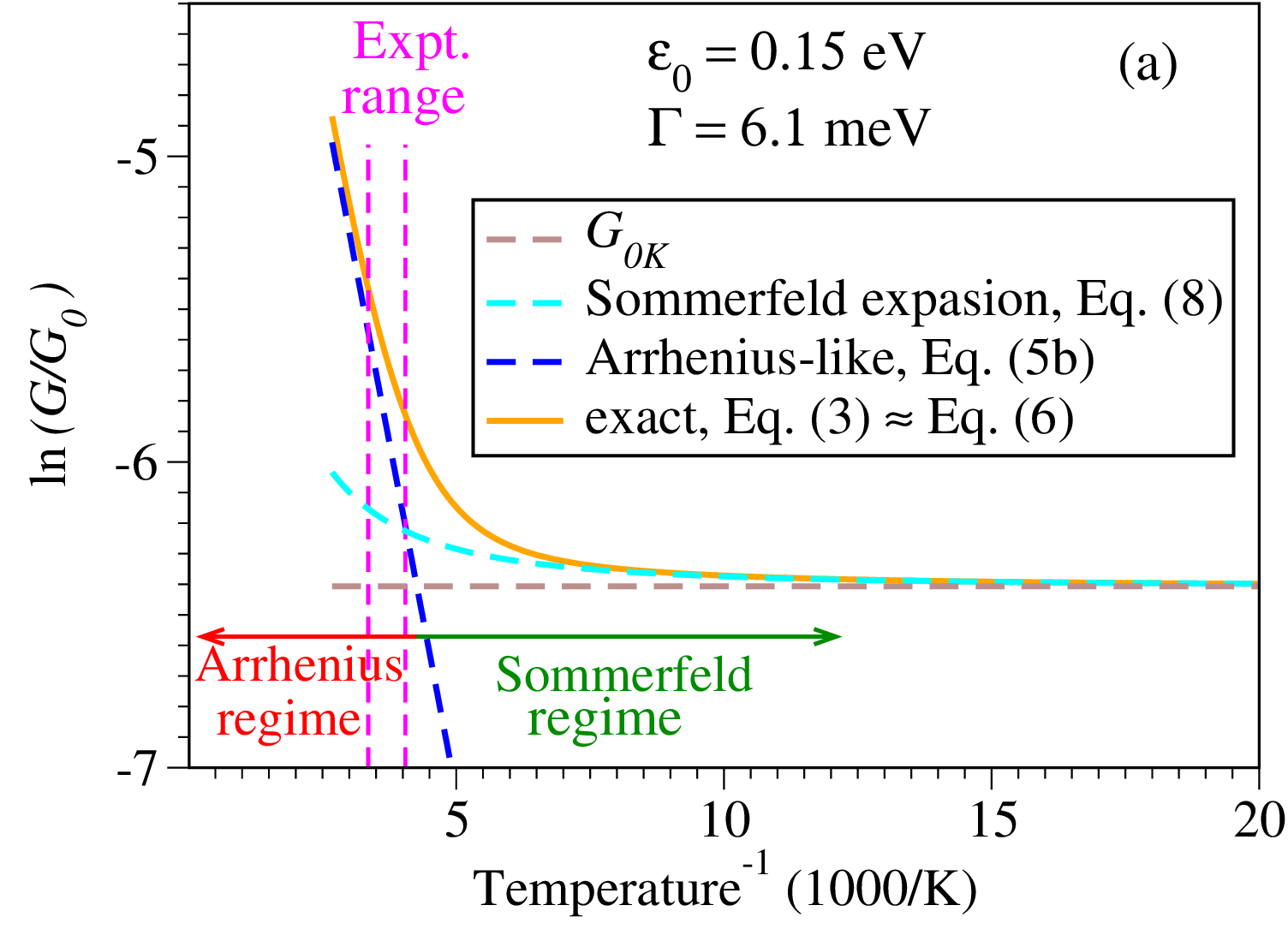}
  } $ $\\[2ex]  
  % plot[0:20] "g_vs_T_Ag_simulation_Delta_0.06.dat" us 4:($6/$10) w l, "g_vs_T_Ag_simulation_Delta_0.006.dat" us 4:($6/$10) w l, "g_vs_T_Ag_simulation_Delta_0.0006.dat" us 4:($6/$10) w l, "g_vs_T_Ag_simulation_Delta_0.000060.dat" us 4:($6/$10) w l
    \centerline{
    \includegraphics[width=0.45\textwidth,angle=0]{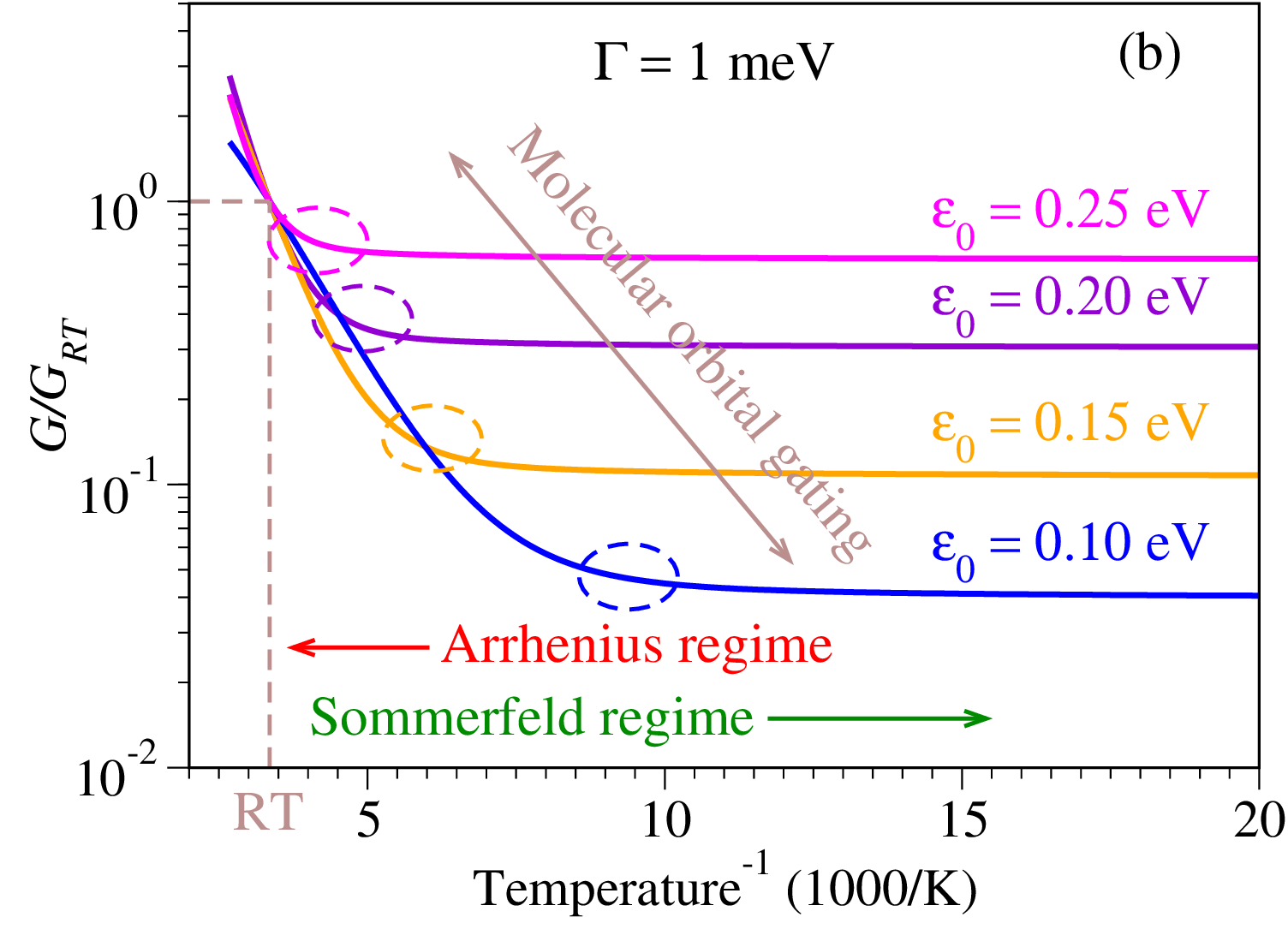}
 } $ $\\[3ex]
  % plot[2:20] "g_vs_T_eta_-0.20_Delta_0.001_w_pccp.dat" us 4:($6/$10) w l, "" us 4:($12/$10) w l, "g_vs_T_eta_-0.15_Delta_0.001_w_pccp.dat" us 4:($6/$10) w l, "" us 4:($12/$10) w l, "g_vs_T_eta_-0.10_Delta_0.001_w_pccp.dat"us 4:($6/$10) w l, "" us 4:($12/$10) w l, "g_vs_T_eta_-0.25_Delta_0.001_w_pccp.dat"  us 4:($6/$10) w l, "" us 4:($12/$10) w l
    \centerline{
    \includegraphics[width=0.45\textwidth,angle=0]{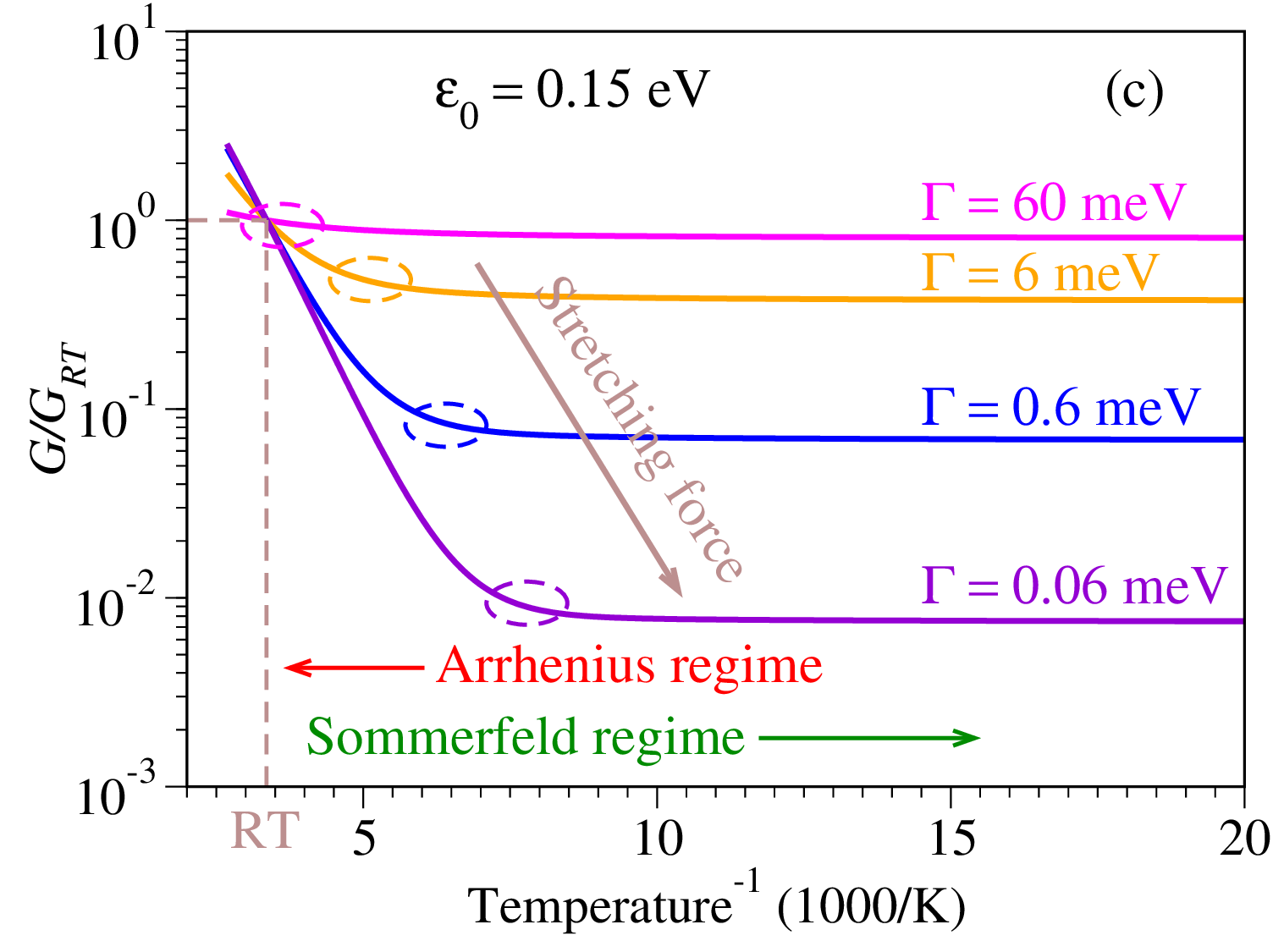}}
% 
  % plot "g_vs_T_Ag_simulation_DeltaRT_sommerfeld.dat" us 2:($12/$8)
  % plot "g_vs_T_Ag_simulation_DeltaPiRT_sommerfeld.dat" us 2:($12/$8)
% 
  \caption{(a)
    Results for conductance obtained using parameters characterizing molecular junctions based on perylene diimide (PDI)
    \cite{Baldea:2018a} beyond the range (indicated by (magenta) vertical lines)
    sampled in measurements \cite{Baldea:2018a} indicate the occurrence of a gradual Arrhenius-Sommerfeld transition
    at experimentally accessible temperatures ($T \sim 170$\,K). 
    (b, c) Results showing how the Sommerfeld-Arrhenius transition (depicted by dashed ellipses)
    can be tuned in junctions subject to a mechanical stretching force or via molecular orbital gating.
    }
  \label{fig:pdi}
\end{figure}

Using parameter values extracted from data measured for
n-type perylene diimide (PDI) molecular junctions with isocyanide surface linkers 
estimated earlier \cite{Baldea:2018a}, further important predictions of the presently deduced \gl~(\ref{eq-g-exact})
are collected in \figname\ref{fig:pdi}.

\figname\ref{fig:pdi}a reveals a crossover
from a thermally activated conductance
exhibiting an Arrhenius-type nearly exponential variation
with $1/T$ at higher temperatures to a nearly $T$-independent $G$ at lower temperatures.
The high-$T$ limit is approximately described by \gl~(\ref{eq-g-arrhenius}).
For the approximate description
of the low temperature range, the Sommerfeld expansion \cite{Sommerfeld:33,AshcroftMermin} can be used.
The expansion of the transmission function around $\varepsilon = 0$ in the RHS of \gl~(\ref{eq-g})
yields in this (Sommerfeld) regime the following expression valid to order $\mathcal{O}\left(T^4\right)$
\begin{eqnarray}
 \frac{G}{G_0} & \simeq & \frac{\Gamma_{g}^{2}}{\varepsilon_0^2 + \Gamma_{a}^2} \left[
   1 + \frac{\varepsilon_0^2 - \Gamma_{a}^{2}/3}{\left(\varepsilon_0^2 + \Gamma_{a}^2\right)^2}  \left(\pi k_B T\right)^2 \right ]
 \nonumber \\ & \xlongrightarrow{\Gamma_{a} \ll \left\vert\varepsilon_0\right\vert} &
 \frac{\Gamma_{g}^{2}}{\varepsilon_0^2} \left(1 + \frac{\pi^2 k_B^2 T^2}{\varepsilon_0^2}\right)
   \label{eq-sommerfeld}
\end{eqnarray}

The inspection of \figname\ref{fig:pdi}a reveals that for the specific junctions envisaged
an Arrhenius-Sommerfeld crossover
should be observable at ``reasonable'' temperatures ($T \sim 170$\,K). Although outside 
the narrow range explored in actual measurements \cite{Baldea:2018a}
($247\,\mbox{K} < T < 298\,\mbox{K}$, cf.~vertical lines in \figname\ref{fig:pdi}a),
the prediction of such moderately low transition temperatures should encourage efforts
for experimental observation.

Importantly, the present results for the tunneling conductance also predict that the
location of the Arrhenius-Sommerfeld transition can be tuned.
Two possibilities to achieve tunability are indicated in \figsname\ref{fig:pdi}: 
by tuning the MO energy $\varepsilon_0$ (\figname\ref{fig:pdi}b) and by tuning the MO-electrode couplings
$\Gamma_{s,t}$ (\figname\ref{fig:pdi}c).
The former can be practically realized via molecular orbital gating \cite{Tao:06,Wandlowski:08,Reed:09}.
For the latter, molecular junctions subject to mechanical deformation are very appealing.
As shown recently \cite{Baldea:2017e,Baldea:2021d},   
the MO-electrode couplings are highly sensitive to an applied stretching force; $\Gamma$'s are primarily
responsible for the observed conductance variations on orders of magnitude \cite{Baldea:2017e,Baldea:2021d}.
This route to unravel an Arrhenius-Sommerfeld transition
may also be relevant for the transport through artificial nanoarrays
\cite{Collier:97,Goldhaber-GordonNature:98,Goldhaber-GordonPRL:98,Markovich:98,Baldea:2002}
wherein the exchange integrals $\tau$'s on which the MO-electrode couplings $\Gamma$'s
depend (cf.~\gl~(\ref{eq-Gamma})) can be tuned in broad ranges.

\ib{Because, after all, the thermal transition between the Sommerfeld and Arrhenius-like regimes discussed above is a theoretical prediction,
  we do not want to end this section before justifying why, above, we have implicitly suggested  
  N-type perylene diimide (PDI) molecular junctions with isocyanide surface linkers
subject to mechanical stretching as promising candidates for the practical realization of a
tunable Arrhenius-Sommerfeld transition.

Reference to PDI-based junctions within the present analysis entirely based on the single-step scenario
was not accidental for three reasons:
\begin{enumerate}
\item First, because it was demonstrated that in these junctions
  the impact of the contact metallurgy elaborated experimentally in Frisbie's group
  on the activation energy safely rules out a two-step hopping scenario \cite{Baldea:2018a};
\item Second, because the foregoing analysis unraveled
that in these junctions the temperature range (around $\sim 170$\,K) wherein the Sommerfeld-Arrhenius transition is expected 
should pose no special experimental problems;
\item Third, because studying PDI-based junctions at variable stretching force should be experimentally feasible.
  Molecular junctions under mechanical deformation is routine technique \cite{Baldea:2017e,Baldea:2019a,Baldea:2019d,Baldea:2021d}
  and topic of continuing interest in the same
  Frisbie's group where the transport through PDI-based junctions at variable temperature to which we referred
  in the present analysis was recorded.
\end{enumerate}
}
\section{Conclusion}
\label{sec:conclusion}
In closing, the present paper emphasizes that a two-step hopping transport mechanism should by no means taken for granted merely because
$G$ versus $1/T$ data follow an Arrhenius pattern.

The exact analytic formula for the single-step tunneling conductance $G$ deduced in this work
demonstrates that molecular junctions wherein charge transport proceeds via 
single-step tunneling can undergo a tunable transition from an
Arrhenius-type strongly temperature dependent regime to a weakly temperature dependent Sommerfeld regime.

Based on this formula, we predicted that molecular junctions subject to variable stretching force or
molecular orbital gating represent possible realizations of the tunable Sommerfeld-Arrhenius transition
and hope to encourage accompanying experimental efforts in this direction.
\section*{Acknowledgments}
Financial support from the German Research Foundation
(DFG Grant No. BA 1799/3-2) in the initial stage of this work and computational support by the
state of Baden-W\"urttemberg through bwHPC and the German Research Foundation through
Grant No.~INST 40/575-1 FUGG (bwUniCluster 2.0, bwForCluster/MLS\&WISO 2.0, and JUSTUS 2.0 cluster) are gratefully acknowledged.
% 
% 
% 
% begin \bibliography{arxiv}

% end \bibliography{arxiv}
% 
% 
% 
% 
\end{document}